\documentclass[prl,a4paper]{revtex4}

\usepackage{amsmath,amssymb}
\usepackage{epsfig}

\newcommand{\numero}[1]{\noindent #1.~}
\newcommand{\la}{\lambda}
\newcommand{\be}{\beta}
\newcommand{\pa}{\partial}
\newcommand{\f}{\frac}
\newcommand{\eqn}{\begin{eqnarray}}
\newcommand{\eqnx}{\end{eqnarray}}
\newcommand{\beg}{\begin{equation}}
\newcommand{\bi}{\begin{itemize}}
\newcommand{\ee}{\end{equation}}
\newcommand{\ei}{\end{itemize}}
\newcommand{\ii}{\item}
\newcommand{\g}{\gamma}

\newcommand{\Ga}{\Gamma}
\newcommand{\de}{\delta}

\newcommand{\nn}{{\cal N}}
\newcommand{\YY}{{\YY}}
\newcommand{\qqqq}{\quad\quad\quad\quad}

\newcommand{\Dl}{\Delta}

\newcommand{\rsq}{{\mathfrak R^2}}

\def\ab{{\alpha_s N_c}/{\pi}}

\def\kb{\bar k^2}
\def\ka{\kappa}

\begin{document}

\title{{\bf QCD, Spin-Glass and String} \\
\vspace{1cm}
\footnotesize{Cracow School of Theoretical Physics, XLVI Course, 2006, 
Zakopane}}

\author{Robi Peschanski}
\email{pesch@spht.saclay.cea.fr}
\affiliation{Service de physique th{\'e}orique, CEA/Saclay,
  91191 Gif-sur-Yvette cedex, France\footnote{%
URA 2306, unit\'e de recherche associ\'ee au CNRS.}}

\begin{abstract}
When quarks and gluons tend to form a dense medium, like in high energy or/and 
heavy-ion collisions, it is interesting to  ask the question which are the 
relevant degrees of freedom that Quantum Chromodynamics predict. The present 
notes correspond to two  lectures given at Zakopane in the (rainy) 
summer 
of 2006, where this question is adressed concretely in two cases, one in the QCD 
regime of weak coupling, the other one at strong coupling. Each case corresponds 
  to the study of an elusive but dynamically important 
transient phase of quarks and gluons expected to appear from Quantum 
Chromodynamics during high energy collisions.

\bigskip

In lecture I, we examine the dynamical phase space of gluon transverse momenta 
near the 
so-called ``saturation'' phase including its fluctuation pattern. ``Saturation'' 
is 
expected to appear when the density of gluons emitted during the collision 
reaches the 
limit when recombination effects cannot  be neglected, even in the 
perturbative QCD 
regime. We demonstrate that the gluon-momenta  exhibit a  nontrivial clustering 
structure, analoguous to ``hot spots'', whose distributions are derived using an 
interesting matching with the 
thermodynamics of directed polymers on a tree with disorder and  its  
``spin-glass'' 
phase.

\bigskip

In lecture II, we turn towards the non-perturbative regime of QCD, which is 
supposed to 
be relevant for the description of the transient phase as quark-gluon plasma  
formed 
during 
heavy-ion collisions at very high energies. Since there is not yet an  available 
field-theoretical scheme for non perturbative QCD in those conditions, we study 
the dynamics of  
strongly 
interacting gauge-theory matter
(modelling quark-gluon plasma)  using the AdS/CFT duality between gauge field 
theory at 
strong coupling and a gravitational background in Anti-de Sitter space. The 
relevant 
gauge theory is a-priori equipped with ${\cal N}=4$ supersymmetries, but  
qualitative 
results may give lessons on this  issue. As an explicit example, we 
show that 
perfect fluid hydrodynamics emerges at large times as the unique nonsingular 
asymptotic 
solution of the nonlinear Einstein equations in the bulk. The gravity dual can 
be 
interpreted as a black hole moving off in the fifth dimension.

\section{Contents}

\section{Lecture I: QCD near Saturation: a Spin-Glass structure}

\vspace{.1cm}  {\bf \numero{1}}{Introduction}

\vspace{.1cm}  {\bf \numero{2}}{Rapidity evolution of QCD dipoles} 

\vspace{.1cm}  {\bf \numero{3}}{Mapping  to thermodynamics of directed polymers}

\vspace{.1cm}  {\bf \numero{4}}{The Spin-Glass phase of gluons}

\vspace{.1cm}  {\bf \numero{5}}{Summary of lecture I}

\section{Lecture II: A Perfect Fluid from String/Gauge Duality}

\vspace{.1cm}  {\bf \numero{6}}{Introduction}

\vspace{.1cm}  {\bf \numero{7}}{AdS/CFT correspondence}

\vspace{.1cm}  {\bf \numero{8}}{Bjorken hydrodynamics}

\vspace{.1cm}  {\bf \numero{9}}{Boost-invariant geometries and Black Holes}

\vspace{.1cm}  {\bf \numero{10}}{Summary of Lecture II}

\end{abstract}

\maketitle

\section{Lecture I: QCD near Saturation: a Spin-Glass structure}

\subsection{Introduction}
\vspace{.1cm}  {\bf \numero{1}} {\it Saturation} in QCD is expected to occur 
when 
parton densities inside an hadronic target are so high that their wave-functions 
overlap. This is expected from the rapidity $Y=\log (W^2)$ evolution of 
deep-inelastic 
scattering amplitudes governed by the Balitsky~Fadin~Kuraev~Lipatov (BFKL) 
kernel 
\cite{bfkl}. The BFKL evolution equation is such that the number of gluons of 
fixed size increases exponentially and would lead without modification to a
violation of unitarity. By contrast, the renormalisation-group evolution 
equations 
following 
Dokshitzer, Gribov and  Lipatov, Altarelli and Parisi (DGLAP) \cite {dglap} 
explains 
the evolution at fixed $Y$ as a function of the hard scale $Q^2.$ It leads to a 
dilute 
system of asymptotically free partons. As schematized in Fig.1, the transition 
to 
saturation  \cite{GLR,CGC} is characterized by a typical 
transverse momentum scale $Q_s(Y)$, depending on the overal rapidity of the 
reaction, when the unitarity bound is reached by the BFKL evolution of the 
amplitude. The two-dimensional plot showing the two QCD evolution schemes and 
the 
transition  boundary to saturation are represented in Fig.1.

\begin{figure} [hb]
\epsfig{file=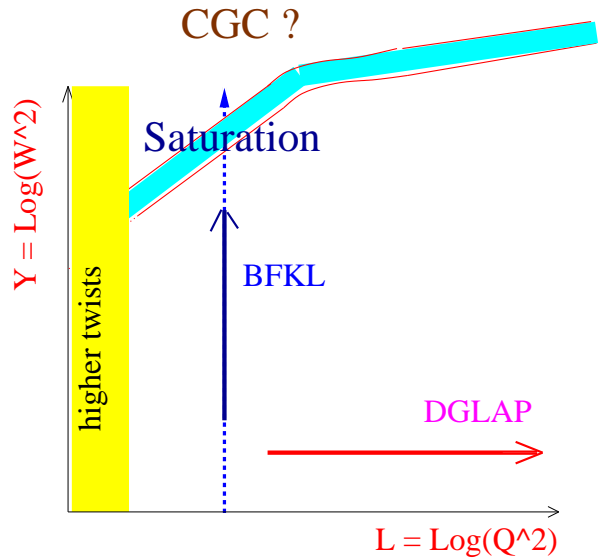,width=8cm}
\caption{{\it Schematic view of the transition region to saturation.} The DGLAP 
and 
BFKL evolution ranges are displayed, together with the saturation region where 
the 
density bounds are reached.}\end{figure}

The problem we  address here is the 
characterization of the gluon-momentum distribution {\it near saturation}. 
Our aim  is to  understand the transverse-momenta spectrum of  the gluons which 
are 
generated by the 
BFKL 
evolution in rapidity, i.e. as characterising a transient QCD phase structure   
near 
saturation {\it as a whole}. The new material contained  in this lecture comes 
from Ref. \cite{RP}.

As a guide for the further developments, the basic structure underlying the 
transition to saturation can be understood in 
terms 
of {\it traveling waves}. If at 
first one neglects the fluctuations (in the  mean-field approximation), the 
effect of saturation on a dipole-target amplitude is described by the nonlinear 
Balitsky-Kovchegov \cite{Balitsky} (BK) equation, where a nonlinear damping term 
 adds to the BFKL equation. As shown in \cite{munier}, this 
equation falls into the universality class of the Fisher and 
Kolmogorov~Petrovsky~Piscounov (F-KPP) nonlinear equation \cite{KPP} which
admits asymptotic traveling-wave solutions. The exponential behaviour of the 
BFKL  
evolution quickly enhances the effects of the tail towards a region where 
finally the nonlinear damping regulates both the traveling-wave propagation and 
structure.

Indeed, one of the major recent challenges in QCD saturation is the problem of  
taking into account the r\^ole of fluctuations, i.e. the structure of gluon 
momenta 
beyond the average. In these conditions it was realized for traveling waves 
\cite{brunet} and thus in the QCD case \cite{Iancu:2004es}, that the 
fluctuations may 
have a surprisingly large effect on the overall solution of 
the nonlinear equations of saturation. Indeed, a fluctuation in the dilute 
regime 
may grow exponentially and thus modify its contribution to the overall 
amplitude. 
Hence, in order to enlarge our understanding of  the QCD evolution with 
rapidity, 
it seems important to give a quantitative description of the pattern of momenta
generated by the BFKL evolution equations for the set of cascading dipoles (or, 
equivalently gluons) {\it near saturation}, which is the subject of the lecture. 

Technically speaking,   we shall 
work 
in the leading order in 
$1/N_c,$ where the QCD dipole framework is valid 
\cite{Mueller:1993rr}. Moreover we will use the diffusive 
approximation 
of the 1-dimensional 
BFKL kernel. In fact, the phase structure  appears quite rich already within 
this 
approximation scheme. Many aspects we will obtain show ``universality'' features 
and 
thus 
are expected to be valid beyond the approximations.

\subsection{Rapidity evolution of  cascading QCD dipoles}

{\bf \numero{2}} Let us 
start 
by briefly describing the QCD evolution of the dipole distributions 
\cite{Mueller:1993rr,salam,Hatta:2006hs}. 

\begin{figure} [ht]
\begin{center}
\epsfig{file=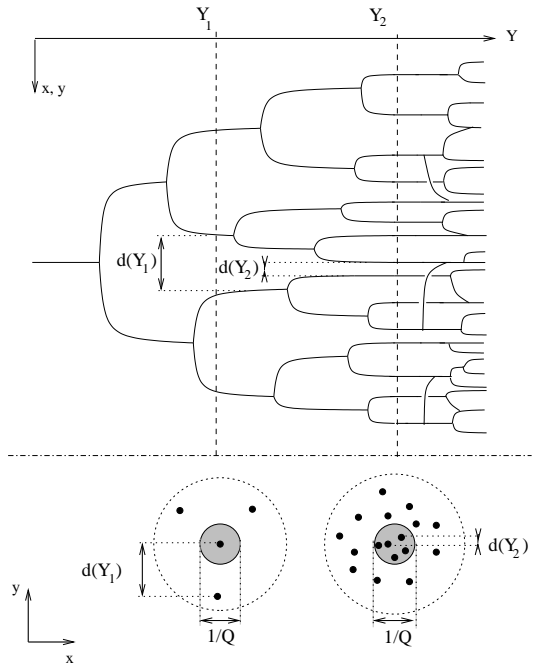,width=7cm}
\end{center}
\caption{{\it BFKL cascading and saturation}.
The  QCD branching process in the BFKL regime and beyond is represented  
along the rapidity axis (upper part). Its 
2-dimensional counterpart in transverse position space is displayed at two 
different rapidities (lower part). The interaction region is represented by a 
shaded disk of size $1/Q$. At rapidity $Y_1$, the interaction still probes 
individual 
dipoles (or gluons), which corresponds to the  exponential  BFKL regime. There 
exists 
a smooth transition to a regime where  the interaction only probes 
groups of dipoles or 
gluons, e.g. at rapidity $Y_2$. This gives a  description of the {\it 
near-saturation} region corresponding to a mean-field approximation, where 
correlations can be neglected. Further in rapidity, $Y> Y_2,$ other dynamical 
effects, such that merging and correlations appear.}
\label{fig2}
\end{figure}

%From Eq.\eqref{equation}, we will mainly retain its physical meaning. 
The 
 structure of BFKL cascading describes a 
2-dimensional tree structure of dipoles in transverse position space evolving 
with 
rapidity. 
Let us, for instance, focus on the rapidity 
evolution starting from one massive $q\bar q$ pair or onium 
\cite{Mueller:1993rr}, see Fig.\ref{fig2}. 
At each 
branching vertex, the wave function of the onium-projectile is described by a 
collection of color dipoles. The dipoles split with a probability per unit of 
rapidity defined by the BFKL kernel \cite{bfkl}
\beg
{\cal K}({\bf v},{\bf w};{\bf z}) = \ab \ \f {\left({\bf v}-{\bf 
w}\right)^2}{\left({\bf v}-{\bf z}\right)^2 \left({\bf z}-{\bf w}\right)^2}\ 
\label{Kernel}
\ee 
describing the dissociation vertex of one dipole 
$({\bf v},{\bf w})$ into two dipoles at 
$({\bf v},{\bf z})$ and $({\bf z},{\bf w}),$ where ${\bf v},{\bf w},{\bf z}$ are 
arbitrary 2-dimensional transverse space coordinates.

As an approximation of  the 2-dimensional formulation of the BFKL kernel 
\eqref{Kernel} obtained when 
one neglects the impact-parameter dependence, we shall 
restrict our analysis in the present paper  to the 1-dimensional reduction of 
the problem to the transverse-momenta moduli $k_i$ of the cascading gluons.  
After Fourier transforming to 
transverse-momentum space, the leading-order BFKL kernel \cite{bfkl} defining 
the rapidity evolution in the 1-dimensional approximation is known 
\cite{Balitsky} to act in transverse momentum space as a  
differential operator of infinite order 
\beg
\chi(-\partial_l) \equiv  2\psi(1)-\psi(-\partial_l)-\psi(1+\partial_l)
\label{eq:lkernel}
\end{equation}
where $l=\log k^2$ and $Y$ is the rapidity in units of the fixed  coupling 
constant $\ab$.

In the sequel, we shall restrict further our analysis to the diffusive 
approximation of 
the BFKL kernel. We thus expand the BFKL kernel  to second order around some 
value $\gamma_c$
\begin{equation}
\chi(\gamma) \sim \chi_c + 
\chi'_c(\gamma-\gamma_c)+{\scriptstyle \frac{1}{2}}\chi''_c(\gamma-\gamma_c)^2
             = A_0 - A_1\gamma + A_2\gamma^2\ ,
\label{eq:coefsai}
\end{equation}
where $\gamma_c$ will be defined in such a way to be relevant for the {\it 
near-saturation} region of the BFKL regime.

Within this diffusive approximation, it is easy to realize that the first term 
($A_0$) is responsible for the exponential increase of the BFKL regime while 
the third term ($A_2$) is a typical diffusion term. The second term ($A_1$) is 
a ``shift'' term since it amounts to  a rapidity-dependent redefinition of the 
kinematic 
variables, as we shall see.
 
In Eq.\eqref{eq:coefsai},
$\gamma_c$ is chosen in order to ensure the validity of the kernel 
\eqref{eq:coefsai} in the transition region from the BFKL regime towards 
saturation.
Indeed, the derivation of asymptotic solutions of the BK equation 
\cite{munier} leads to consider  the 
condition 
\begin{equation}
\chi(\gamma_c) = \gamma_c\ \chi'(\gamma_c)\ 
\label{eq:condition}
\end{equation}
whose solution determines $\gamma_c.$

This condition applied to the kernel formula \eqref{eq:lkernel} gives $\gamma_c 
=\sqrt {A_0/ A_2}\approx 0.6275...$ and  
$\{A_0,A_1,A_2\}\approx 
\{9.55,25.56,24.26\}.$ These numbers may appear anecdotic, but they fully 
characterize the critical parameters of the saturation transition, as we will 
realize later on.
For different kernels, e.g. including next-leading log effects 
\cite{Peschanski:2005ic}, they could be different, of course.  But, then the 
traveling-wave solution will be in a different ``universality class''  in 
mathematical terms.

\subsection{Mapping  to thermodynamics of directed polymers}

{\bf \numero{3}}From the properties  in transverse-momentum space and within the 
1-dimensional diffusive approximation (\ref{eq:coefsai}),  we already noticed 
that the BFKL kernel models boils down to a branching, shift  and diffusion  
operator acting in the gluon transverse-momentum-squared space. Hence the 
cascade of gluons  can be put in correspondence with  a continuous branching,  
velocity-shift and  diffusion  probabilistic process, see Fig.\ref{fig2}, whose 
probability by unit 
of rapidity is defined by the coefficients $A_i$ of \eqref{eq:coefsai}.

\begin{figure}[ht]
\epsfig{file=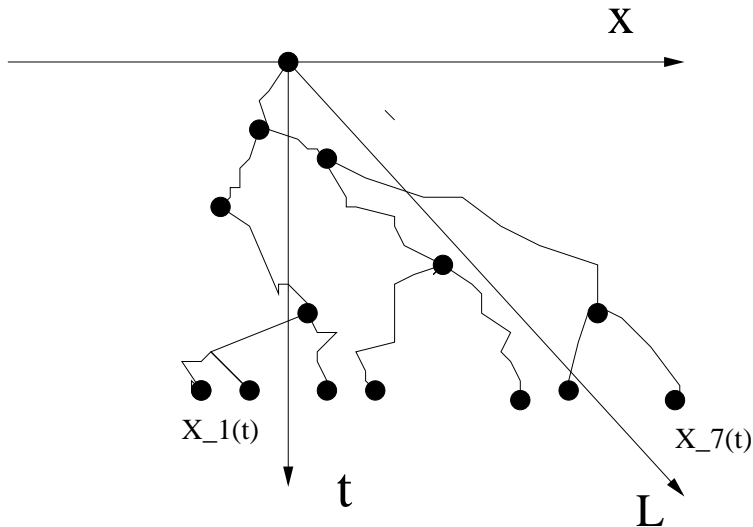,width=10cm}
\caption{{\it Branching diffusion model for polymers}. The coordinates 
$x_1(t)\cdots x_7(t)$ correspond to the random paths along the tree in the 
$x,t$ phase-space. The oblique axis is for $L= \be x +{A_1}/{A_0}\ t,$ which 
takes into account the ``time-drift'' in the mapping to the QCD problem.}
\label{fig:fkpp}
\end{figure}
 
Let us first introduce the notion of  gluon-momenta ``histories'' $k_i(y).$ They  
register the evolution of the gluon-momenta starting from the unique initial 
gluon momentum $k(0)$ and terminating with the specific $i$-th momentum $k_i,$ 
after successive branchings. They  define  a random function of the {\it running 
rapidity}  $y,$ with $0 \le y \le 
Y,$  
the final 
rapidity range when the evolution ends up (say, for a given total energy). It is 
obvious that two different histories $k_i(y)$ and $k_j(y)$ are equal before the 
rapidity when they branch away from their  common ancestor.

In order to fromulate precisely the mapping to the polymer problem, we then 
introduce random paths $x_i(t)$ using formal {\it space} and  {\it 
time} coordinates $x, t\ ,\ 0\le t \le {\Delta t} $  which we  relate to  
gluon-momenta ``histories'' as follows:
\begin{equation}
 \ y = \ \f t{A_0}\ ;\quad 
\quad \log  {k_i^2(y)} \equiv   - \be\ (x_i(t)-x(0)) + ({A_0}-{A_1}) y
\label{eq:mapping}
\end{equation}
where $(A_0-A_1)y$ is a conveniently chosen and deterministic ``drift term'', 
$x(0)$ is an arbitrarily fixed origin of an unique initial gluon and thus the 
same for all subsequent random paths. The random paths $x_i(t)$ are 
generated by a continuous
branching and Brownian diffusion process in space-time (cf. Fig.\ref{fig:fkpp}).

As we shall determine later on, the important parameter $\be,$ which plays the 
r\^ole of an inverse of the temperature $T$ for the Brownian movements of the 
polymer process, is given by
\begin{equation}
\f 1T \equiv \be = \sqrt {{2A_2}/{A_0}}\ .
\label{eq:mu}
\end{equation}

In fact, the relation \eqref{eq:mu}  will be required by the condition that the 
stochastic process of  random paths describes the BFKL regime of QCD {\it near 
saturation}. Another choice of $\be$ would eventually describe the same 
branching 
process but in other conditions. Hence the condition \eqref{eq:mu} will be 
crucial to determine the QCD phase at saturation (within the diffusive 
approximation).

Let  now introduce the tree-by-tree random function defined as the partition 
function of the random paths system
\begin{equation}
Z(t) \equiv\sum_{i=1}^{n}e^{-\be x_i(t)} = e^{-\be x_0+A_1y}\times \f 1n 
\sum_{i=1}^{n=e^{A_0 y}} k_i^2(y)\ \propto \  e^{A_1 y}\times \kb(y)\ ,
\label{Z}
\ee
where $\f 1n\sum_{i=1}^{n} k_i^2(y)\equiv \kb(y)$ is the  {\it event-by-event} 
average over gluon momenta at rapidity $y.$ Note that one has to distinguish 
 $\ \overline{\ \cdots\ }$ i.e. the average made over only {\it one} event from  
 $\langle\  \cdots \ 
\rangle,$
which 
denotes  the average over samples (or events). 

$Z(t)$ is an event-by-event random function. The physical properties are 
obtained 
by averaging various observables over the events. Note that the distinction 
between averaging over one event and the  sample-to-sample averaging appears 
naturally   in the statistical physics problem in terms of ``quenched'' 
disorder: 
the time scale associated with the  averaging over one random tree structure is  
much shorter than the one corresponding to the averaging over random
trees. 

With these definitions,  $Z$ appears to be  nothing else than  the partition 
function for 
the 
model of directed polymers on a random tree  \cite{Derrida}. 

Let us now justify the 
connection of the 
directed-polymer 
properties with the description of the gluon-momentum phase {\it near  
saturation} by rederiving the known saturation features from the statistical 
model 
point-of-view. 
Using the known properties \cite{Derrida} of the partition function of the 
polymer problem, one finds
\beg
\log Q_s^2 \equiv \langle\  \log \kb \ \rangle \equiv \langle\  \log Z \ \rangle 
-A_1 Y=  
\left[(2\sqrt{ {A_2}{A_0}}- A_1) Y-\f 32\sqrt{\f {A_0}{A_2}}\log Y\right] +  
{\cal O}(1)\ ,
\label{eq:saturation}
\ee
which in fact matches exactly  the asymptotic expansion found in \cite{munier} 
for 
the 
saturation scale.

In the same way, the solution of the statistical-physics problem allows to 
derive the event-by-event spectrum of 
the {\it free energy} $\log Z (t)$ of the system around its average. From  
\eqref{Z}, one gets 
\begin{equation}
\nn(\kb,Y) \sim {\cal P}\left(\log Z-\langle\log Z\rangle\right) \ \propto \ 
\log\left[ \f {\kb}{Q_s ^2}\right]\ 
\exp\left\{-\sqrt{\f {A_0}{A_2}} \log \left[\f {\kb}{Q_s ^2}\right]\right\}\ 
,
\label{eq:Proba2}
\end{equation}
which is the well-known geometrical scaling property, empirically found in 
Ref.\cite{geom} and theoretically derived  in  \cite{munier} from
the BK equation for the dipole amplitude $\nn(\kb,Y).$

Both properties \eqref{eq:saturation},\eqref{eq:Proba2} prove the consistency of 
the model with the properties expected from saturation. We shall then look for 
other properties of the cascading gluon model. It is important to realize that 
this consistency 
fails for a different  choice of the parameter $\be$ different from 
\eqref{eq:mu}. 
This justifies 
a-posteriori the identification of the equivalent temperature of the system in 
the 
gluon/polymer mapping framework.

\subsection{The Spin-Glass Phase of gluons}
\vspace{.2cm}  {\bf \numero{4}}Let us now come to the main new results 
concerning the 
determination of structure of the gluon-momentum phase at saturation.

The striking property of the directed polymer problem on a random tree is the 
spin-glass structure of the low temperature phase. As we shall see this will 
translate directly into a specific  {\it clustering} structure  of gluon 
transverse momenta in their phase near the ``unitarity limit'', see 
Fig.\ref{valleys}.
\begin{figure} [ht]
\begin{center}
\epsfig{file=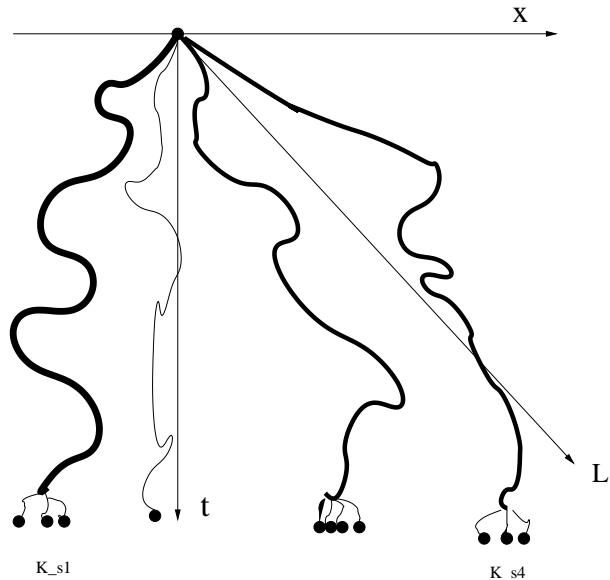,width=8cm}
\end{center}
\caption{{\it The clustering structure of gluons near saturation}. The drawing 
represents the $s_1\cdots 
s_4$ clusters near momenta $k_{s1}\cdots k_{s4}$. They branch either near $t\ll 
1$ or 
$(\Delta t-t)\ll 1,$ where $\Delta t$ is the total amount of time evolution.}
\label{valleys}
\end{figure}

 Following the   
relation \eqref{eq:mu} , one is led to 
consider the polymer system at temperature 
$T$ with
\beg
\f {T_c-T}{T_c} \equiv 1-\f {\be_c}{\be} = 1- \g_c\ ,
\label{eq:T}
\ee
where $\g_c$ is the critical exponent defined by \eqref{eq:condition}. We are 
thus naturally led to consider the 
low-temperature phase ($T<T_c$), at some distance $T_c-T$ from the critical 
temperature 
$T_c=\scriptstyle{1/{\sqrt 2}}.$ In the language of traveling waves \cite{wave}, 
this corresponds to a ``pulled-front'' condition with ``frozen''and  
``universal'' velocity  and front profile.

As derived in \cite{Derrida}, the 
phase space of the polymer problem is structured in ``valleys'' which are in the 
same universality class as those of the Random Energy Model (REM) \cite{REM} and 
of the infinite range Sherrington-Kirkpatrick (SK) model \cite{SK}. 

Translating these results in terms of gluon-momenta moduli, the phase 
space 
landscape consists in event-by-event distribution of clusters of momenta around 
some values $k_{si}^2 \equiv  1/(n_i)\ \sum_{i\in si} k_{i}^2,$ where $n_i$ is 
the cluster multiplicity. The probability weights to find a cluster $s_i$ after 
the whole evolution range $Y$ is 
defined by
\beg
W_{si}= \f {\sum_{i\in si} k_{i}^2}{\sum_i k_i^2},
\label{y}
\ee
where the summation in the numerator is over the momenta of gluons within the 
$s_i^{th}$ cluster (see Fig.\ref{valleys}). The normalized distribution of 
weights 
$W_{si}$ thus allows one to study  the  probability distribution of clusters. 
The clustering tree structure, (called ``ultrametric'' in statistical mechanics) 
is 
the most prominent feature of spin-glass systems
\cite{overlap}.

Note again that, for the QCD problem, this property is proved for momenta in the 
region of the ``unitarity limit'', or more concretely in the momentum region 
around the saturation scale. This means that the cluster average-momentum  is 
also such that $k_{si}^2 = {\cal O}(Q^2_s).$ Hence the clustering structure is 
expected to appear in the range  which belongs to the traveling-wave 
front \cite{munier} or, equivalently, of clustering with finite fluctuations 
around  the saturation scale. 

In order to quantify the cluster structure, one may introduce  a well-known 
``overlap function''    in statistical 
physics of 
spin-glasses
\cite{overlap}. Translating the definitions (cf. \cite{Derrida}) in terms of the 
QCD problem, one  introduces an event-by-event indicator of the 
strength of clustering which is built in from the weights \eqref{y}, namely
${\cal Y}=\sum_{si} W^2_{si}.$
The non-trivial probability distribution of overlaps ${\Pi}({\cal Y})$ possesses 
some universality features, since 
it is identical to the 
one of the REM and SK models and shares many qualitative similarities with 
other systems possessing a spin-glass phase \cite{toulouse,toulouse1}. Exemples 
are  given in 
Fig.\ref{fig:overlap}.
\begin{figure}[t]
\epsfig{file=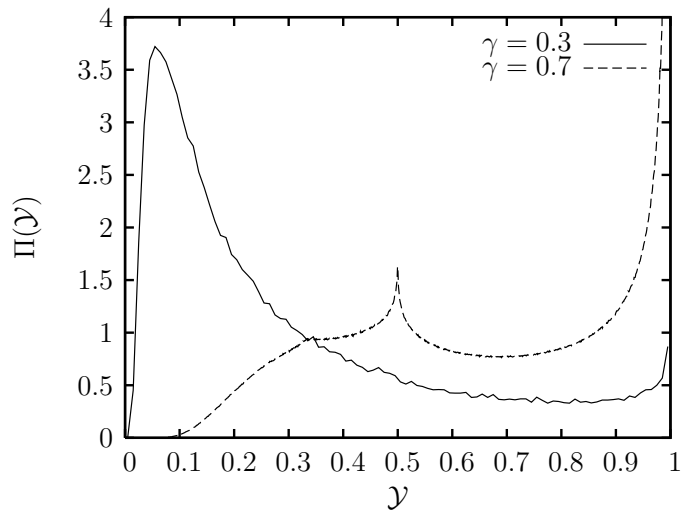,width=9cm}
\caption{{\it The probability distribution of overlaps ${ \Pi}({\cal Y})$.} The 
figure (the simulation is by courtesy from  \cite{private},  using the method of 
Ref.\cite{toulouse1}) is drawn both for  the theoretical QCD
value 
$\g_c = 0.7,$ and for $\g_c = 0.3$ for comparison. The statistics used for the 
simulation is $10^8$ events in $10^3$ bins for $\g_c = 0.3$ and $3.25\ 10^6$ 
events in $10^2$ bins for $\g_c = 0.7.$}
\label{fig:overlap}
\end{figure}

The rather involved probability distribution $\Pi({\cal Y})$   is quite 
intringuing. It 
possesses a priori an infinite number of singularities at 
${\cal Y } = 1/n, n\ {\rm integer}.$ It can be seen when the temperature 
is significantly lower from the critical value, see for instance the curve for 
$1-T/T_c = .7$ in Fig.\ref{fig:overlap}. However, the predicted curve for the 
QCD value $1-T/T_c \sim .3$ is smoother and shows only a final cusp at $W_s=1$ 
within the considered statistics.
It thus seems that configurations with  only one  cluster 
can be  
more prominent than the otherwise smooth generic landscape. However, it is also 
a 
``fuzzy'' landscape since many clusters of various sizes seem to coexist in 
general.

\subsection{Summary of lecture I}
\vspace{.1cm}  {\bf \numero{5}}We 
investigated the landscape of transverse momenta in gluon cascading around the 
saturation scale at asymptotic rapidity. Limiting our study to a diffusive 
1-dimensional modelization of the BFKL regime of gluon cascading, we make use of 
a mapping on a statistical physics model for directed polymers propagating along 
random tree structures at fixed temperature. We then  focus our study  on the 
region near the unitarity limit where information can be obtained on saturation, 
at least in the mean-field approximation. Our main result is to  find a 
low-temperature spin-glass structure of phase space, characterized by 
event-by-event clustering of gluon transverse momenta (in modulus) in the 
vicinity of the rapidity-dependent saturation scale. The weight  distribution of 
clusters and the probability of  momenta overlap during the rapidity evolution 
are derived.

Interestingly enough the clusters at asymptotic rapidity are branching either 
near the beginning (``overlap $0$'' or $y/Y \ll 1$) or near the end (``overlap 
$1$'' or $1-y/Y \ll 1$) of the cascading event. The probability distribution of 
overlaps is derived and shows a rich singularity structure.

On a phenomenological ground, it is remarkable that saturation density effects 
are not equally spread out on the event-by-event set of gluons; our study 
suggests that there exists random spots of higher density whose distribution may 
possess some universality properties. In fact it is natural to  expect this 
clustering property to be present not only in momentum modulus (as we could  
demonstrate) but also in momentum azimuth-angle. This is reminiscent   of the 
``hot spots'' which were some time  ago \cite{Bartels:1992bx} advocated from the 
production of forward jets in deep-inelastic scattering at high energy 
(small-$x$). The observability of the cluster  distribution through the 
properties of ``hot spots'' is an 
interesting 
possibility.

\eject
\section{Lecture II: A Perfect Fluid from Gauge/Gravity Duality}

\subsection{Introduction}

{\bf \numero{6}}From the first years of the running of heavy-ion collisions at 
RHIC,
evidence has been advocated  that various observables are in good agreement
with models based on hydrodynamics \cite{hydro} and with
quark-gluon plasma (QGP) in a strongly coupled regime \cite{strongQGP}. 
To a large extent it seems that the QGP behaves approximately as a
perfect fluid as was first considered in \cite{Bjorken}. A schematic view of the 
theoretical expectations in given in Fig.\ref{QGP}.
It is a challenge in QCD to derive from first principles the
properties of the dynamics of 
a strongly interacting plasma formed in heavy-ion collisions and in
particular to understand why the perfect-fluid hydrodynamic equations
appear to be relevant.

\begin{figure}[ht]
\epsfig{file=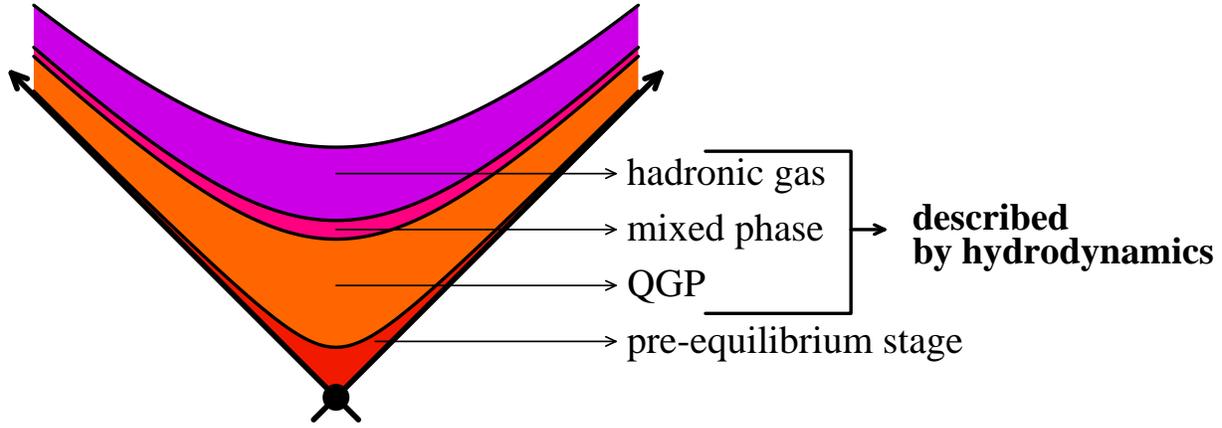,width=16cm}
\caption{{\it Scenario for the quark-gluon plasma(QGP) formation}. After a 
pre-equilibrium stage in a heavy-ion collision, probably governed by a 
weak-coupling but dense regime, the QGP is formed with local equilibrium and 
hydrodynamic properties.}
\label{QGP}
\end{figure} 

Even if the experimental situation is still developing and rather
complex, it is worth simplifying the problem in order to
be able to attack it with appropriate theoretical tools.
Recently the AdS/CFT correspondence \cite{adscft,adscftrev} emerged as  a new
approach to 
study strongly coupled gauge theories. This has been largely worked
out in the supersymmetric case and in particular for the conformal
case of $\nn=4$ super Yang-Mills theory (SYM). Interestingly enough, since
the QGP is a deconfined and strongly interacting phase of QCD we could expect
that results for the nonconfining $\nn=4$ theory may be relevant or at least 
informative on the unknown strong coupling QCD problem. We
will start from  this assumption in our work.

In this lecture we focus on the spacetime evolution of the gauge theory
(4d) energy-momentum tensor, and derive its asymptotic behaviour from
the solutions of the nonlinear Einstein equations of the gravity
dual. 

Imposing the absence of curvature singularities in the gravity dual,
we will show that, in the boost invariant setting (as in
\cite{Bjorken}), perfect fluid hydrodynamics emerges from the AdS/CFT
solution at large times. The new material contained here comes  from Refs. 
\cite{janik}.

%The plan of our paper is as follows. In section 2 we review the
%Bjorken hydrodynamics on the gauge theory side. Then, in section 3, we
%setup a general framework of deriving a gravity dual for a given
%energy-momentum tensor on the boundary, based on the
%holographic renormalization method. 
%In section 4 we derive the
%large proper-time behaviour of the boost-invariant gravity duals by solving
%analytically the corresponding nonlinear Einstein equations in the bulk. 
%In section 5 we arrive at the physical solution by requiring the absence
%of curvature singularities. This constraint selects perfect fluid
%hydrodynamics in the 4D gauge theory. We close the paper with
%conclusions and outlook.
 
\subsection{String/Gauge fields Duality}

{\bf \numero{7}}As an introduction to our lecture, let us briefly recall some 
aspects of the String/Gauge Duality. The AdS/CFT correspondence 
\cite{adscft}  has many interesting formal and 
physical facets. Concerning 
the 
aspects which are of interest for our problem, it allows one to find 
relations 
between gauge field theories at strong coupling and string gravity at 
weak 
coupling  in the limit of large number of colours ($N_c\!\to\! \infty$). It can 
be examined  quite precisely in  the 
AdS$_5$/CFT$_4$ 
case which conformal field theory  corresponds to $SU(N)$ gauge 
theory 
with ${\cal N} \!=\!4$ supersymmetries.

Let us  recall the canonical derivation leading to  the AdS$_5$ 
background , 
see 
Fig.\ref{ads}. One starts from  the (super)gravity classical 
solution of a system of $N\ D_3$-branes in a $10\!-\!D$ space of the (type 
IIB) 
superstrings. 
The metrics solution of the (super)Einstein equations read
\beg
\label{super}
ds^2=f^{\!-\!1/2} (\!-\!dt^2\!+\!\sum_{1\!-\!3}dx_i^2) 
\!+\!f^{1/2}(dr^2\!+\!r^2d\Omega_5) 
\ ,
\ee
where the first four coordinates are on the brane and $r$ corresponds to 
the  
coordinate along the normal to the branes. In formula (\ref{super}), one defines
\beg
f=1+\frac {R^4}{r^4}\ ;\ \ \ \ \ R=4\pi g^2_{YM}\alpha'^2 {N} \ ,
\label{R}
\ee
where $g^2_{YM}{N}$ is the `t Hooft-Yang-Mills coupling and $\alpha'$ the 
string 
tension. 
\begin{figure}[ht]
\epsfig{file=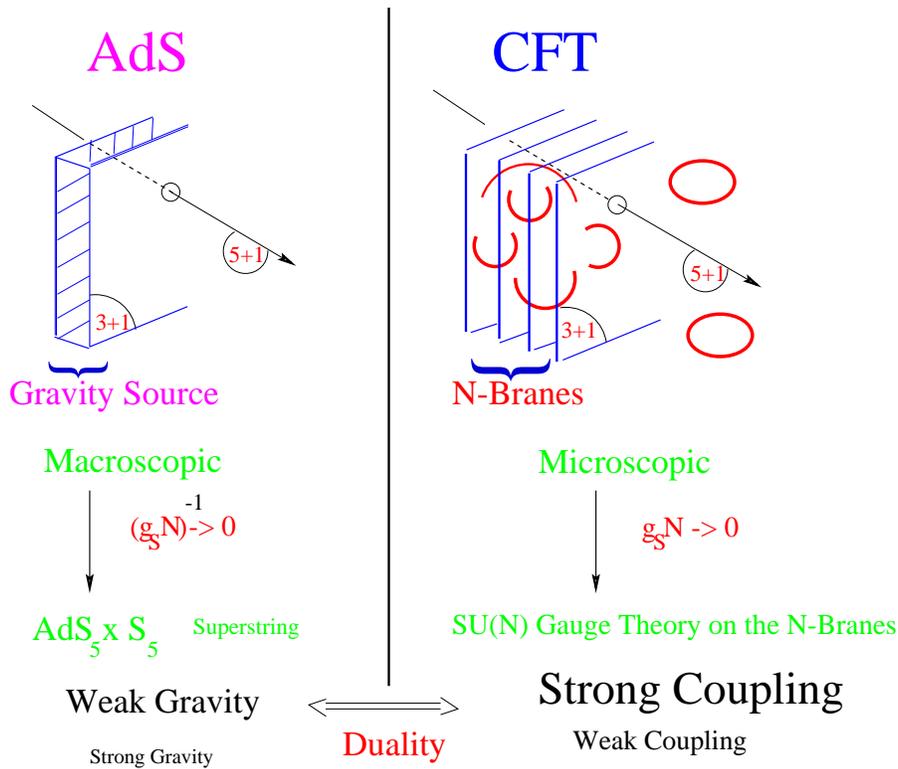,width=12cm}
\caption{{\it Schematic view of the Gauge/String Duality.} Left: The string 
background is the Anti-de Sitter space (AdS); Right: the gauge theory is a 
Conformal field theory  (CFT) on the 4-d N-branes.}
\label{ads}
 \end{figure} 
One considers the   limiting behaviour considered by 
Maldacena, 
where 
one zooms on the  neighbourhood of the branes  while in the same time going to 
the limit of 
weak 
string 
slope $\alpha'.$ The near-by space-time is thus distorted due to the 
(super) 
gravitational field of the branes. One goes to the limit where 
\beg
\ R\ fixed\ ; \ \frac {\alpha'(\to 0)} {r (\to 0)}\to z\ fixed\ .
\ee
This, from the second equation of (\ref{R}) obviously implies 
\beg
{\alpha'\to 0}\ , \ g^2_{YM} 
{N}\sim \f{1}{\alpha'^2}\to \infty \ ,
\ee
{\it i.e.} both a weak coupling  limit for the string theory and   a strong 
coupling limit for the dual gauge field theory.
By reorganizing the two parts of the metrics one obtains
\beg
ds^2={ \frac 1{z^2} (-dt^2+\sum_{1-3}dx_i^2+ dz^2)} +  {R^2 
d\Omega_5}\ ,
\label{AdS}
\ee
which corresponds to the 
{AdS$_5$} $\times \  {S_5}$ background structure,  ${S_5}$ being  
the 5-sphere. More detailed analysis shows that the isometry group of 
the  
5-sphere is the geometrical dual of the ${\cal N}\! =\!4$ 
supersymmetries. More intricate is the quantum number dual to $N_c,$ the number 
of 
colours, which is the invariant charge carried by the Ramond-Ramond form 
field.

\subsection{Bjorken hydrodynamics}

{\bf \numero{8}}Coming back to the physical world, a model of the central 
rapidity region
of heavy-ion reactions based on
hydrodynamics was pioneered in 
\cite{Bjorken} and involved the assumption of boost
invariance. Our goal is to  study the dynamics of strongly interacting
gauge-theory matter assuming boost
invariance.

We will be interested in the spacetime evolution of the
energy-momentum tensor $T_{\mu\nu}$ of the gauge-theory matter. It is
convenient to introduce proper-time ($\tau$) and space-rapidity ($y$)
coordinates in the longitudinal position plane:
\beg
x^0=\tau \cosh y \qqqq x^1=\tau \sinh y \ .
\ee  
In these coordinates, all components of the
energy momentum tensor can be expressed (see \cite{janik}) in terms of a {\em 
single}
function $f(\tau)$: 
\beg
\label{e.tgen}
T_{\mu\nu}\! = \!
\left(\begin{tabular}{cccc}
$f(\tau)$ & 0 &0 & 0 \\
0 & $-\tau^3 \f{d}{d\tau} f(\tau)\!-\!\tau^2 f(\tau)$ & 0 & 0 \\
0 & 0 & $f(\tau)\!+\! \f{1}{2}\tau \f{d}{d\tau} f(\tau)$ & 0 \\
0 & 0 & 0 & $f(\tau)\!+\! \f{1}{2}\tau \f{d}{d\tau} f(\tau)$
\end{tabular}\right)
\ee
where the matrix $T_{\mu\nu}$ is expressed in $(\tau,y,x_1,x_2)$
coordinates. 

Furthermore the function $f(\tau)$ is constrained to verify
\beg
\label{e.enpos}
f(\tau)\geq 0 \qqqq f'(\tau)\leq 0 \qqqq
\tau f'(\tau) \geq -4 f(\tau) \ .
\ee
The dynamics of the gauge
theory should pick a specific $f(\tau)$. A perfect fluid or a fluid with
nonzero viscosity and/or other transport coefficients will lead to
different choices of $f(\tau)$. 

We thus address the problem of
determination of the function $f(\tau)$ from the AdS/CFT correspondence. 
Let us first describe two distinct cases of physical interest:

For a perfect fluid (Bjorken hydrodynamics)
$f(\tau)\sim {1}/{\tau^{\f{4}{3}}},$
while for  a ``free streaming case'' expected just at the beginning of the 
interaction \cite{Kovchegov} ,
$f(\tau)\sim {1}/{\tau}.$ 
In the following we  introduce a family of $f(\tau)$
with the large $\tau$ behaviour of the form
\beg
\label{s}
f(\tau) \sim \tau^{-s}.
\ee

\subsection{Boost-invariant geometries}

{\bf \numero{9}}The most general form of the bulk metric respecting 
boost-invariance can be written
\beg
\label{e.ansatz}
ds^2=\f{-e^{a(\tau,z)} d\tau^2 +\tau^2 e^{b(\tau,z)} dy^2
  +e^{c(\tau,z)} dx^2_\perp}{z^2} +\f{dz^2}{z^2}  \ .
\ee
The three coefficient functions can be (non trivially) derived from the Einstein 
equations
\beg
\label{e.einst}
R_{\mu\nu}-\f{1}{2}g_{\mu\nu} R - 6\, g_{\mu\nu}=0\ ,
\ee
in the asymptotic limit where $\tau \to \infty.$ Interestingly enough, they 
depend only on the scaling variable $v=z/\tau^{s/4},$ where $s$ labels the 
one-parameter  family of solutions \eqref{s}.

After quite painful calculations, the solution reads \cite{janik}
\beg
a(v) = A(v)-2m(v) \
b(v) = A(v)+(2s-2) m(v) \
c(v)= A(v)+(2-s) m(v)
\label{e.c}
\ee
where
\beg
A(v)=\f{1}{2} \left( \log(1+\Dl(s)\,v^4) +\log(1-\Dl(s)\, v^4) \right) \
m(v)=\f{1}{4\Dl(s)} \left( \log(1+\Dl(s)\,v^4) -\log(1-\Dl(s)\, v^4)
\right)
\ee
with
\beg
\Dl(s)=\sqrt{\f{3s^2-8s+8}{24}} \ .
\ee

Specializing first to
the perfect fluid case, this gives rise to the  following {\em asymptotic} 
geometry
\eqn
\label{e.flgeom}
z^2\ ds^2= - \f{\left( 1-\f{e_0}{3}
      \f{z^4}{\tau^{4/3}}\right)^2}{1+\f{e_0}{3}\f{z^4}{\tau^{4/3}}} d\tau^2+
\left( {\textstyle 1+\f{e_0}{3} \f{z^4}{\tau^{4/3}}}\right) (\tau^2
      dy^2 +dx^2_\perp)  + {dz^2}
\eqnx
Remarkably enough this geometry can be identified (in suitable metrics)  to be a 
 {\it moving} Black Hole, which evolves in the fifth dimension $z.$

For the free streaming case, one finds
\beg
z^2\ ds^2={  -(1\!+\!\f{v^4}{\sqrt{8}})^{\f{1\!-\!2\sqrt{2}}{2}}
  (1\!-\!\f{v^4}{\sqrt{8}})^{\f{1\!+\!2\sqrt{2}}{2}} dt^2+ 
 (1\!+\!\f{v^4}{\sqrt{8}})^{\f{1}{2}}
  (1\!-\!\f{v^4}{\sqrt{8}})^{\f{1}{2}} \tau^2 dy^2 + 
(1\!+\!\f{v^4}{\sqrt{8}})^{\f{1\!+\!\sqrt{2}}{2}}
  (1\!-\!\f{v^4}{\sqrt{8}})^{\f{1\!-\!\sqrt{2}}{2}} 
  dx^2_\perp } +{dz^2}\ , 
\ee
which is qualitatively different from the perfect fluid case, in
particular it displays singularities or zeroes at
$v^4=\sqrt{8}$ in all coefficients. 
More generally, it is possible to show \cite{janik} that the perfect fluid case 
is the only one free of physical singularities, namely singularities which 
cannot be removed by a change of coordinates. In order to check this feature, we 
considered the 
metric-invariant curvature scalar 
\beg
\rsq=R^{\mu\nu\alpha\beta}R_{\mu\nu\alpha\beta}\ .
\ee
As an illustration, we represent this property in Fig.\ref{F}, where the value 
of 
$\rsq$ 
is studied as a function of the distance from the horizon, for $s$ values at the 
perfect fluid point  and very near-by values.
\begin{figure}[t]
\epsfig{file=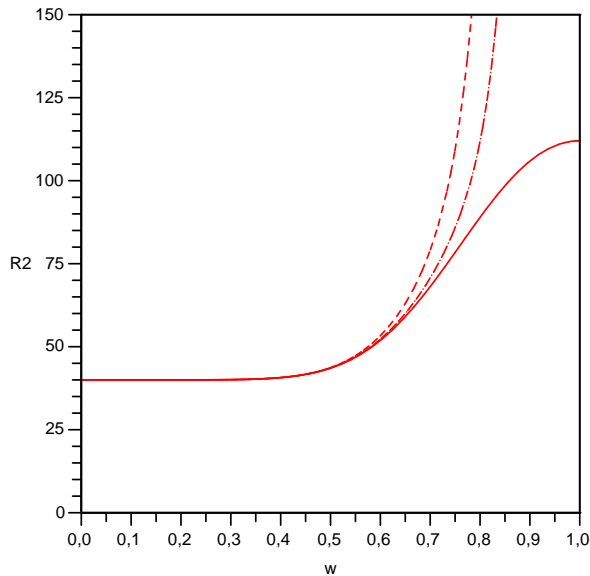,height=9cm}
%\centerline{\includegraphics[height=9cm]{rsq.eps}}
\caption{{\it The curvature scalar $\rsq .$}  $\rsq $ is calculated as a 
function of
  $w=v/\Dl(s)^{1/4}$ for the perfect fluid case $s=4/3$ (solid line),
  $s=4/3-0.1$ (dotted line) and $s=4/3+0.2$ (dashed line).}
\label{F}\end{figure}
Let us add some comments on the specific features of our approach and
results. We concentrate on looking for solutions of the
full nonlinear Einstein equations. It would be interesting to confront
this approach with the linearization methods of refs. \cite{son}. In
particular viscosity terms are expected to appear in the study of
subasymptotic terms. 
Note that the possibility of black hole formation in the {\em dual}
geometry has been argued in
ref. \cite{nastase}. 
More specifically, the geometry of a brane moving w.r.t. a black hole
background has been advocated in ref. \cite{zahedsin} for the dual
description of the cooling and expansion of a quark-gluon
plasma. In our case we could interpret the solution (\ref{e.flgeom})
as a kind of `mirror' situation in terms of a black hole moving off from
the AdS boundary. 

\subsection{Summary of Lecture II}
{\bf \numero{10}} We have introduced a general framework for
studying the dynamics of matter (plasma) in strongly coupled gauge theory
using the AdS/CFT correspondence for the $\nn=4$ SYM theory. We constructed  
dual geometries for given 4-dimensional gauge theory
energy-momentum tensor profiles.  Further imposing boost-invariant dynamics 
inspired by the Bjorken
  hydrodynamic picture, we have found the corresponding asymptotic
  solutions of the nonlinear Einstein equations.  Among the family of asymptotic 
solutions, the only one with
  bounded curvature scalars is the gravity dual of a perfect fluid
  through its energy-momentum tensor profile. This selected nonsingular 
solution, given by the metric
  (\ref{e.flgeom}), 
  corresponds to a black hole moving off in the 5th dimension as a
  function of the physical proper time.  As an application of this 
framework, we 
can obtain \cite{janik} the thermalization time of the perfect fluid, which 
describes the 
decay 
back to equilibrium of a scalar excitation of the perfect fluid out of 
equilibrium, by computation of the quasi-normal modes of the moving Black Hole. 
In 
some sense, the moving Black Hole is a quite {\it stable} geometric 
configuration. 
We  conjecture that it may represent, through the Gauge/Gravity duality, a 
powerful 
``attractor'' for the QGP evolution, or even perhaps for more general evolution 
of a strongly coupled system of quarks and gluons.

\bigskip

\noindent{}{\bf Acknowledgments.} Many aspects depicted in Lecture I come from 
constant collaboration and discussion inside (and outside) our  ``Saturation 
Team'' in Saclay 2006, in particular Edmond Iancu, Cyrille Marquet, Gregory 
Soyez. I warmly thank Romuald Janik for its major contribution in the 
fruitful collaboration whose results are discussed in Lecture II.

\end{document}